\documentclass[final,3p,times,authoryear]{elsarticle}

\usepackage{graphicx}
\usepackage{amsmath}
\usepackage{amssymb}
\usepackage{amsfonts}
\usepackage{mathrsfs}
\usepackage{adjustbox}

\usepackage{mdframed}
\usepackage[figuresright]{rotating}
\usepackage{courier} %makes output with \texttt{} look nicer
\usepackage[final]{changes}
\usepackage{setspace}
\usepackage{dutchcal}

\graphicspath{ {figures/} }

\usepackage[utf8x]{inputenc} %Umlaute

\usepackage{nomencl}
\usepackage{etoolbox}
\usepackage{wrapfig}
\usepackage{stfloats} % having multiple figures
\usepackage{framed} % Framing content
\usepackage[skins,breakable, most]{tcolorbox}

\renewcommand*\nompreamble{\begin{multicols}{2}} %\renewcommand*\nompostamble{\end{multicols}}
\usepackage{epstopdf} %transfers eps into pdf before preparation
\usepackage{todonotes}
\usepackage{setspace}
\usepackage{siunitx}

\usepackage[caption = false]{subfig}
\usepackage[hyphens]{url}
\usepackage{graphicx}
\usepackage{booktabs}

\usepackage{lmodern,textcomp}
\usepackage{rotating} % Rotating table
\usepackage{multicol} % Multiple columns environment
\usepackage{multirow} 
\usepackage{array} % Allows to adjust column and line width
\usepackage{mathtools, cuted}
\usepackage{lscape}
\usepackage{tabularx} % allows defining width of tables
\usepackage{longtable}  %for better-spaced tables over multiple pages, head at top pf every new page
\usepackage[flushleft]{threeparttable}
\usepackage{threeparttablex} %for "ThreePartTable" environment
\usepackage{tabulary}
\usepackage{amssymb}
\usepackage{pifont}% http://ctan.org/pkg/pifont
\RequirePackage[ngerman=ngerman-x-latest]{hyphsubst}
\usepackage{microtype}
\usepackage[american]{babel} 
\usepackage{float}
\usepackage[normalem]{ulem}
\useunder{\uline}{\ul}{}
\usepackage{textcomp}

\biboptions{square,comma,sort&compress}
\usepackage{array}
\newcolumntype{C}[1]{>{\centering\arraybackslash}p{#1}}

\setcitestyle{numbers}
\journal{}
\makeatletter
\def\ps@pprintTitle{%
 \let\@oddhead\@empty
 \let\@evenhead\@empty
 \def\@oddfoot{}%
 \let\@evenfoot\@oddfoot}
\makeatother

\usepackage{longtable}  %for better-spaced tables
\usepackage{tabulary}
\usepackage{pdfpages}

\usepackage[acronym]{glossaries}
\usepackage{glossary-inline}
\renewcommand*\nompreamble{\begin{multicols}{2}}
\makeglossaries
\newacronym{eu}{EU}{European Union} 
\newacronym{susic}{SUSIC}{Smart Utilities and Sustainable Infrastructure Change} 
\newacronym{pv}{PV}{Rooftop Photovoltaic Systems}
\newacronym{eea}{EEA}{Energy Efficient Appliances}
\newacronym{gt}{GT}{Green Electricity Tariffs}
\newacronym{lcts}{LCTs}{Low Carbon Technologies}
\newacronym{lct}{LCT}{Low Carbon Technology}
\newacronym{nep}{NEP}{New Environmental Paradigm}
\newacronym{sem}{SEM}{Structural Equation Modelling}
\newacronym{tpb}{TPB}{Theory of Planned Behaviour}
\newacronym{plssem}{PLS-SEM}{Partial Least Squares Structural Equation Modeling}
\newacronym{cbsem}{CB-SEM}{Covariance-Based Structural Equation Modeling}
\newacronym{vif}{VIF}{Variance Inflation Factor}
\newacronym{gof}{GoF}{Goodness of Fit}
\newacronym[description={Average Variance Extracted}]{ave}{AVE}{Average Variance Extracted}
\newacronym[description={One-Way Analysis of Variance}]{anova}{ANOVA}{One-Way Analysis of Variance}
\newacronym{plsmga}{PLS-MGA}{partial least squares multi-group analysis}
\newacronym{att}{ATT}{Attitude}
\newacronym{int}{INT}{Intention}
\newacronym{novs}{NOVS}{Novelty Seeking}
\newacronym{envc}{ENVC}{Environmental Concern}
\newacronym{socn}{SOCN}{Social Norm}
\newacronym{pbc}{PBC}{Perceived Behavioural Control}
\newacronym{finb}{FINB}{Financial Benefit}
\newacronym{envb}{ENVB}{Environmental Benefit}
\newacronym[description={Alternative Fuel Vehicle}]{afv}{AFV}{Alternative Fuel Vehicle}
\newacronym{doi}{DOI}{Diffusion of Innovation Theory}
\newacronym{vbn}{VBN}{Value-Belief-Norm Theory}
\newacronym{ev}{EV}{electric vehicle}
\newacronym{wtp}{WTP}{willingness to pay}

\newcommand\blfootnote[1]{%
 \begingroup
 \renewcommand\thefootnote{}\footnote{#1}%
 \addtocounter{footnote}{-1}%
 \endgroup
}
\begin{document}
\begin{frontmatter}

%% title
\title{Green or greedy: the relationship between perceived benefits and homeowners' intention to adopt residential low-carbon technologies}

%% \author[institution_abbreviation]{author name}
\author[FHWS,CAE]{Fabian Scheller\corref{cor1}}
\ead{fabian.scheller@thws.de}
\author[DTU]{Karyn Morrissey}
\author[DIW]{Karsten Neuhoff}
\author[DTU]{Dogan Keles}

%% \address[abbreviation]{Institution}
\address[FHWS]{Institute Zero Carbon (IZEC), Technical University of Applied Sciences Würzburg-Schweinfurt (THWS), Ignaz-Schön-Straße 11, Schweinfurt, Germany}
\address[CAE]{Center for Applied Energy Research (CAE), Magdalene-Schoch-Straße 3, 97074 Würzburg, Germany}
\address[DTU]{Climate and Energy Policy, Department of Technology, Management and Economics, Technical University of Denmark (DTU),  Akademivej, 2800 Kgs. Lyngby, Denmark}
\address[DIW]{Climate Policy Department, German Institute for Economic Research (DIW), Mohrenstraße 58, 10117 Berlin, Germany}

%% abstract
\begin{abstract}

Transitioning to a net-zero economy requires a nuanced understanding of homeowners' decision-making pathways when considering the adoption of \acrlong{lcts} (\acrshort{lcts}). These \acrshort{lcts} present both personal and collective benefits, with positive perceptions critically influencing attitudes and intentions. Our study analyses the relationship between two primary benefits: the household-level financial gain and the broader environmental advantage. Focusing on the intention to adopt \acrlong{pv}, \acrlong{eea}, and \acrlong{gt}, we employ \acrlong{plssem} to demonstrate that the adoption intention of the \acrshort{lcts} is underpinned by the \acrlong{tpb}. Attitudes toward the \acrshort{lcts} are more strongly related to product-specific benefits than affective constructs. In terms of evaluative benefits, environmental benefits exhibit a higher positive association with attitude formation compared to financial benefits. However, this relationship switches as homeowners move through the decision process with the financial benefits of selected \acrshort{lcts} having a consistently higher association with adoption intention. At the same time, financial benefits also positively affect attitudes. Observing this trend across both low- and high-cost \acrshort{lcts}, we recommend that policymakers amplify homeowners' recognition of the individual benefits intrinsic to \acrshort{lcts} and enact measures that ensure these financial benefits.

\end{abstract}

%% keywords


%\abbreviations{S1, symbol1; S2, symbol2.}

\end{frontmatter}

\section*{Highlights}
\begin{itemize}
\item Relationship between perceived benefits and attitude and intention to adopt \acrshort{lcts} is analysed.
\item Determinants of \acrshort{tpb} consistently drive the adoption intentions for both low- and high-cost \acrshort{lcts}.
\item Environmental benefits have a greater association with attitude than financial benefits across  \acrshort{lcts}.
\item Intention to adopt a \acrshort{lct} is based on perceived financial rather than environmental benefits.
\item Greater awareness of the future financial benefits of LCTs is required to accelerate diffusion.
\end{itemize}

\blfootnote{\printglossary[type=\acronymtype,style=inline,nonumberlist]}

\section{Introduction}
\label{S:1}
Contributing to up to 70\% of global greenhouse gas emissions, the transition to a net-zero economy necessitates a shift in consumption behaviours. Household energy consumption significantly impacts the residential sector's carbon footprint \citep{dubois2019starts}. Thus, strategies to reduce this consumption are vital to the EU's targets of achieving net-zero emissions by 2050 \citep{creutzig2018towards}. The adoption of a variety of residential \acrshort{lcts} could enhance efficiency. However, despite the rising public concern about climate change and increasingly favourable attitudes towards \acrshort{lcts} among homeowners \citep{Geels.2018}, their widespread adoption remains limited \citep{schleich2019energy}. Economic barriers, such as upfront costs, are believed to hinder the adoption of pricier \acrshort{lct} like \acrlong{pv} (\acrshort{pv}) \citep{KORCAJ2015407}. Yet, the initial costs for other \acrshort{lct} can be minimal or even nil, like switching to a \acrlong{gt} (\acrshort{gt}), or moderate, as in the purchase of efficient household 'white goods' (\acrlong{eea} \acrshort{eea}).

Grasping the factors that shape the decision-making pathway for adopting \acrshort{lcts} is pivotal to hastening the diffusion rate needed to achieve our net-zero objectives \citep{creutzig2018towards}. In addressing this, past research has introduced several theoretical models to comprehend the decision-making process behind homeowners adopting \acrshort{lcts}. Examples include the diffusion of innovations theory \cite{Rogers.2003} and the value belief norm theory of pro-environmental behaviour \citep{Stern.2000}. Nevertheless, the most prevalent remains the \acrlong{tpb} (\acrshort{tpb}) \citep{Ajzen.1991}. Since the \acrshort{tpb} elucidates adoption behaviour, it logically extends to the adoption of \acrshort{lcts}, which is a consumer behaviour.

The \acrshort{tpb} suggests that behaviours arise from prior intentions. This intention encapsulates motivational elements that impact specific behaviours \citep{Ajzen.1991}. A robust link between intention and behaviour is posited, meaning stronger intentions typically lead to the actual performance of a behaviour \citep{ajzen2020}. At its core, the model assumes that individuals usually act based on their intentions \citep{Klockner.2015}. Factors such as attitude towards the behaviour, perceived social pressures (subjective norms), and belief in one's capability (perceived behavioural control) influence these intentions \citep{Ajzen.1991}. For instance, attitudes are shaped by beliefs about the outcomes of a behaviour and how one evaluates those outcomes \citep{ajzen2007}. Moreover, underlying these intentions are specific criteria that individuals use to gauge the benefits of a behaviour \citep{schulte2022meta,Wolske.2017}. Such behaviours are adopted when individuals perceive advantages, be it monetary gains or social prestige \citep{KORCAJ2015407}.

Adopting \acrshort{lcts} entails a behavioural shift among private, single-family homeowners. Leveraging the \acrshort{tpb}, research has shown that positive attitudes towards \acrshort{lcts} lead to stronger intentions to embrace these technologies \citep{Wolske.2017}. Thus, pinpointing the evaluative criteria that underpin (i) the homeowner's attitudes and (ii) subsequent intentions to adopt an \acrshort{lct} is paramount in understanding the changes throughout the decision-making process. In a \acrshort{pv} study, Korcaj et al. \cite{KORCAJ2015407} underscored that underlying motives of evaluative benefits can either be collective (e.g., environmental and economic), benefitting the community, or individual (e.g., autarky, financial, social status), exclusively benefitting the homeowner. Notably, environmental benefits (from 'green' decision-makers) and financial benefits (from 'greedy' decision-makers) are frequently cited concerning collective and individual benefits \citep{Abreu.2019}. Further \acrshort{tpb} research on \acrshort{lcts} has pointed out that the selection and relative significance of evaluative criteria may shift as decision-making progresses \citep{schulte2022meta}. For instance, studies found that as people transition from forming attitudes to crafting intentions, perceived financial benefits show a more substantial link with the intention to adopt \acrshort{pv} than environmental benefits do \citep{Claudy.2013,Wolske.2017}.

Drawing on the findings of a recent meta-analysis regarding adoption intention by Schulte et al. \cite{schulte2022meta}, we employ an extended version of the \acrshort{tpb} (as depicted in Figure \ref{fig:overviewTPB}) to delve into the influence of perceived environmental and financial benefits on attitudes towards, and the subsequent intention to adopt, three \acrshort{lcts} spanning various cost points. Specifically, the chosen \acrshort{lcts} include \acrshort{pv} (high-cost), \acrshort{eea} (medium-cost), and \acrshort{gt} (low-cost). Our hypothesis posits that although environmental benefits will be significant in shaping attitudes towards all three \acrshort{lcts}, the intention to adopt \acrshort{lcts} will predominantly correlate with financial benefits over environmental ones. In this scenario, as decision-makers approach the final adoption decision, individual benefits tend to wield greater influence than collective ones. By concentrating exclusively on the preliminary two stages of decision-making and targeting homeowners who haven not yet adopted the discussed \acrshort{lct}, our study sidesteps potential complications like recall bias—a common issue when respondents are asked to recollect their motivations for opting for a previously adopted product or service.

Our contributions to the extant literature are manifold. 
\begin{enumerate}
    \item We extend the existing research by examining the role of perceived personal and collective benefits on attitudes towards, and the subsequent intention to adopt, three \acrshort{lcts}. Our focus centers on determining the associations as decision-makers progress through their decision-making journey. We underscore the significance of product-specific benefits, contrasting them with affective constructs, in shaping homeowners' attitudes and intentions towards \acrshort{lcts}.
    
   \item By employing uniform survey instruments and focusing on consistent variables and constructs for each \acrshort{lct}, we offer an improved understanding of how individual and collective benefits function across three \acrshort{lcts} with varying cost spectrums. Such an approach grants us a robust framework for comparing across the \acrshort{lcts}, setting our research apart from earlier studies that largely fixated on a single \acrshort{lct}.
   
   \item In line with prior research, our data reaffirms that aspects like gender, income, and age remain peripheral in influencing attitudes and intentions related to \acrshort{lcts}. Our  multi-group analysis across different socio-demographic groups strengthens this assertion, underlining the universal relevance and transferability of our findings.
\end{enumerate}

%Our contribution to the previous research is threefold. First, this study adds to the current research base on the role of perceived personal and collective benefits on attitudes to, and subsequent intention to adopt three \acrshort{lcts}. The focus is on understanding the associations as decision makers transition through the decision-making process. We highlight the importance of product-specific benefits rather than affective constructs in shaping homeowners' attitudes and intention to adopt \acrshort{lcts}. Second, by utilising a similar survey instrument and querying homogeneous variables for each of the \acrshort{lcts}, we gain insights into the relative role of individual and collective benefits for three \acrshort{lcts} with different price points. This approach allows for enhanced comparability across the \acrshort{lcts} compared to previous studies that primarily focused on a single \acrshort{lct}. Third, our analysis reinforces previous research that gender, income, and age do not significantly influence attitudes and intentions towards \acrshort{lcts}. Thereby, we provide additional evidence through a multi-group analysis across different socio-demographic sub-groups, highlighting the generalisability of the results.

\section{Conceptual model and hypotheses}
\label{S:2}

\subsection{Extended theoretical framework}
\begin{figure}[htp]
	\centering
	\captionsetup{width=0.95\linewidth}
 \includegraphics[width=0.85\textwidth]{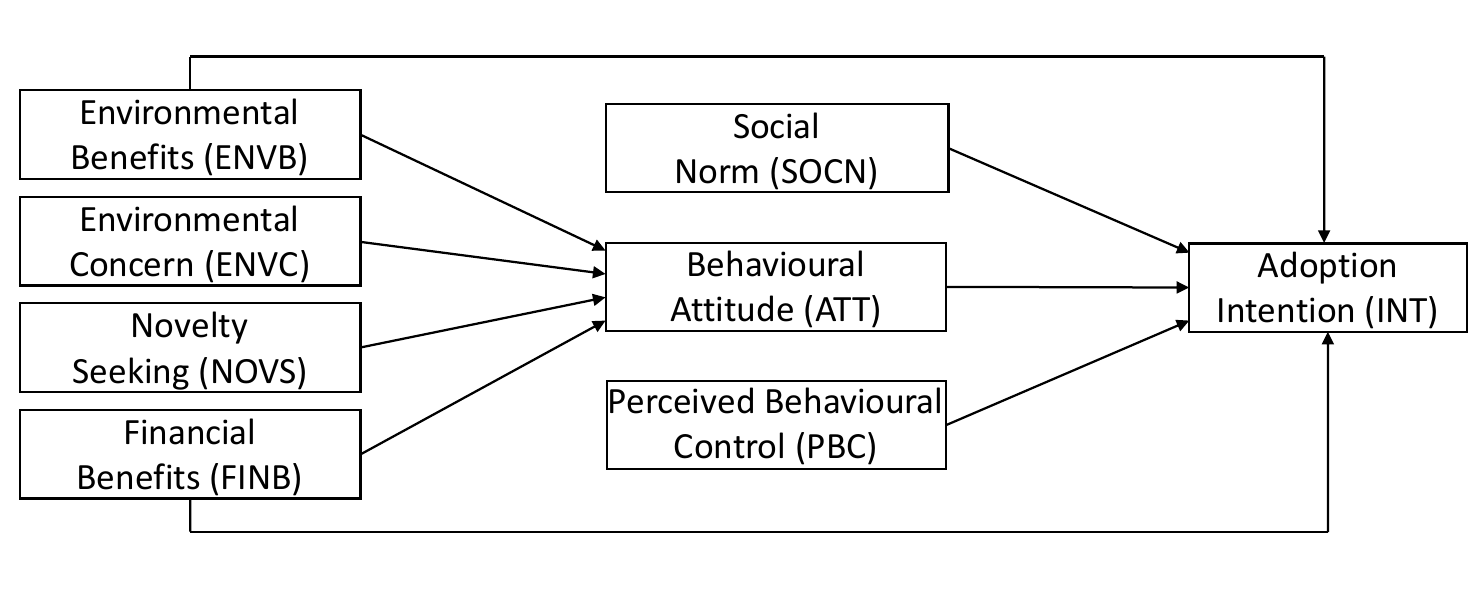}
	\caption[Research model and hypotheses]{Extended \acrshort{tpb} research model focusing on the relationship between perceived environmental or financial benefits and the attitude towards or intention to adopt \acrshort{lcts}. The abbreviations in the brackets describe the names of the variables used in our analysis.}
	\label{fig:overviewTPB}
\end{figure}

The \acrshort{tpb} is one of the most widely used behaviour prediction models in psychological research \citep{KORCAJ2015407,schulte2022meta}. To understand the role of individual versus collective benefits underpinning homeowners' attitudes towards, and intention to adopt \acrshort{lcts}, we expand the \acrshort{tpb} by adding two evaluative constructs. Perceived environmental and financial benefits were chosen as they demonstrated a stronger association with both attitude towards and intention to adopt \acrshort{lcts} than other queried benefits \citep{Abreu.2019}. People are often perceived as selfish when prioritising their personal benefits over the environmental improvements. We also included two affective constructs, novelty seeking and environmental concern, for attitude formation. In relation to the addition of the two affective constructs, it was deemed vital to understand participants' general level of environmental concern, as the questions regarding environmental benefits were developed to understand the perceived environmental benefits of each \acrshort{lct} \citep{bamberg2003does,chen2014developing}. Novelty seeking was also included since previous research found that those with a higher affinity for novelty tend to be more interested in adopting \acrshort{lcts} \citep{Chen2014,Jansson2011,Wolske.2017}. An overview of the conceptual model of this study is outlined in Figure \ref{fig:overviewTPB}.

\subsection{Hypotheses formation}
Concerning the relative importance of individual and collective benefits throughout the decision-making process, research demonstrated that positive environmental beliefs, involvement or knowledge, warm glow benefits and utilitarian environmental benefits positively influence attitudes to \acrshort{gt} \citep{paladino2019black} and \acrshort{eea} \citep{waris2020empirical,bhutto2020adoption,ha2012predicting}. However, for more costly \acrshort{lcts} such as \acrshort{pv}, financial benefits showed a stronger link with positive attitudes towards these technologies \citep{KORCAJ2015407, Wolske.2017}. Since empirical studies have consistently found a positive relationship between collectivism and green attitudes in the context of environmentally-friendly attitudes \citep{leonidou2010antecedents,mccarty2001influence}, we posit our first hypothesis as:\\

\noindent\textit{H1: Perceived environmental benefits of each of the respective  \acrshort{lct} have a stronger association with a positive attitude towards adopting the  \acrshort{lct} than perceived financial benefits at the population level.}\\

Subsequent studies, however, found that as individuals progress through the decision-making process, transitioning from attitudes to intention, personal benefits are more tightly linked to the intention to adopt \acrshort{pv} \citep{Claudy.2013,Wolske.2017}. Personal benefits such as financial factors were also considered more vital than environmental benefits \cite{Arroyo2019,BERGEK2017547,Wolske.2017}. Concurrently, one cannot overlook the positive influence of environmental benefits. Different studies have illustrated a positive relationship between intention and environmental awareness or norms \citep{Mundaca2020,Wolske.2020,Vasseur2015}.

Ecological benefits were found to have a positive association with the intention to adopt \acrshort{eea} \cite{paparoidamis2019making}. When including financial and environmental benefits, however, perceived financial benefits had a higher coefficient \cite{fatoki2020factors}. A further study in Germany focusing on consumers with low energy demand indicated that they are not deterred about the charges of \acrshort{gt} \citep{gerpott2010determinants}. Ozaki \cite{Ozaki} found that the personal benefits of switching to \acrshort{gt} mitigated the inconvenience of switching contracts. Other publications demonstrated the relevance of benefits but did not clearly distinguish between personal or collective benefits \citep{AzizN.S..2017,hua2019antecedents}. Based on this research and distinguishing between personal and collective benefits and their relative role in intention to adopt \acrshort{lcts}, we hypothesise that:
\\

\noindent\textit{H2: Perceived financial benefits of each of the respective  \acrshort{lct} show a stronger association with intention to adopt the \acrshort{lct} than perceived environmental benefits at the population level.}\\

Lastly, prior research has indicated that socio-demographics do not modify the intention to adopt innovations \citep{Arts.2011} or \acrshort{lcts} \citep{Kastner.2015,Klockner.2013,diamantopoulos2003can}. For instance, studies by Paparoidamis et al. \cite{paparoidamis2019making}and Hansla et al. \cite{Hansla.2008} did not observe that age, gender, or income had a significant relationship with intention to adopt \acrshort{eea} \cite{gerpott2010determinants}. The same was observed for income and adoption intention in a German survey \cite{arkesteijn2005early}. However, the landscape is more intricate for \acrshort{pv}, with some studies pinpointing the influence of income or wealth \citep{Best.2019,Jan.2020,Nair.2010} and age \citep{Zander.2020,bondio2018technology}. Consequently, our final hypothesis examines the potential moderating effects of socio-demographic factors:\\

\noindent\textit{H3: The relationships observed in H1 and H2 remain valid across socio-demographic sub-groups income, age, and gender for different \acrshort{lcts}.}\\

\section{Methodology}
\label{S:3}

\subsection{Survey design and data acquisition}
\label{S:3.1}
A quantitative survey was used to analyse and compare variables influencing the adoption intention of \acrshort{pv}, \acrshort{eea}, and \acrshort{gt} in Germany. Whilst the questions were asked in German, an English translation of the questions and items is provided in Table \ref{tab: operational_determ}. Each of the three questionnaires consisted of three sections.

The survey began with a series of screening questions. Only respondents who indicated that they owned a house (single-family home or duplex) and who were solely or jointly responsible for household investment decisions were invited to proceed. These respondents were then asked if they had already adopted \acrshort{pv}, \acrshort{eea}, and/or \acrshort{gt}. As individuals tend to assess their own decisions positively once they have been made \cite{Rai.2016} and to assess the intention to adopt unambiguously, respondents were not permitted to possess the \acrshort{lct} in the focus of their survey. However, they were allowed to own one or both of the other \acrshort{lcts}. Only respondents aged 18 to 75 were admitted. % Respondents below the age of 18 were excluded from the survey due to a lack of legal capacity. In contrast, the maximum age was set at 75 to ensure comparability between results.

In the main part of the survey, the questions related to the constructs of the extended \acrshort{tpb} as depicted in Figure \ref{fig:overviewTPB}. An outline of the items is provided in Table \ref{tab: operational_determ}. Respondents were first asked about their purchase intention to adopt either \acrshort{pv}, \acrshort{eea}, or \acrshort{gt}. Subsequently, constructs for adoption intention were queried using multiple manifest (observable) variables. Established scales were available for the constructs of novelty seeking (NOVS) proposed by Manning et al. \cite{Manning1995}, and environmental concern (ENVC) developed by Dunlap et al. \cite{Dunlap2000} and recommended by Best \cite{Best2011}. For this survey questionnaire, each item was measured using a seven-point Likert response format. New scales with the same response format were created for the constructs of attitude (ATT), social norm (SOCN) and financial (FINB) and environmental benefits (ENVB). Following the recommendation by \cite{Fishbein.2010}, these scales comprised three to six items per construct as outlined in Table \ref{tab: operational_determ}. %The attitude towards \acrshort{pv}, \acrshort{eea}, and \acrshort{gt} was measured using four items on self-directed attitude and action of interest as suggested by \cite{Ajzen.1977}. According to the review of \cite{FARROW20171}, the social norm had different conceptualisations, including descriptive and injunctive norms. Whilst the descriptive norm refers to what most people do, the injunctive norm refers to what most people approve of doing \cite{FARROW20171}. In line with \cite{Fishbein.2010}, a scale for the social norm that encompasses both conceptualisations was included for each of the \acrshort{lct}. For the construct of financial evaluation, four items corresponding to the individual reflection on the financial viability of \acrshort{pv}, \acrshort{eea}, and \acrshort{gt} adoption were also based on the questions \citep{schwarz2007umweltinnovationen}.

Additionally, barriers were examined in terms of perceived behavioural control (PBC) which might influence the level of intention. Here too, each item was measured using a seven-point Likert response format if the barrier was relevant for the respondent ("a little" (1) to "very much" (7)). However, respondents also had the option to select "not applicable" for any barriers that weren't relevant ("not applicable" (0)). In line with various suggestions from Ajzen \cite{ajzen2020}, and modified items from Claudy et al, Rai et al, and Wolske et al. \cite{Wolske2017,Claudy.2013,Rai.2016} as shown in Table \ref{tab: operational_determ}, the survey included controlling factors related to skills and abilities, availability of the \acrshort{lcts} in the area, necessary time investment, affordability issues, and other suitability constraints of the respondent's home.

\newpage
\scriptsize
\renewcommand{\arraystretch}{1.2}
\setlength{\tabcolsep}{5pt}
\begin{longtable}{p{.2\textwidth} p{.1\textwidth} p{.62\textwidth}}
\caption{Constructs and associated items of the three surveys. While the items of the constructs novelty seeking, independent judgment making, and environmental concern were the same in all of the surveys regardless of the \acrshort{lct} in focus, the constructs attitude, social norm, financial evaluation, and environmental evaluation were similar but dependent on the \acrshort{lct} in focus (\acrshort{pv}, \acrshort{eea}, or \acrshort{gt}). The same applied to the intention variable which was dependent on the \acrshort{lct}.}
\label{tab: operational_determ}\\
\textbf{Construct} & \textbf{Item}  & \textbf{Content}  \\
\hline
\endhead
Intention (INT) & int & How strong is your current intention to adopt [a photovoltaic system $\mid$ an energy efficient fridge $\mid$ a green electricity tariff] for your private home?     \\ \hline
 & novs\_1 & “I often seek out information about new products and brands."   \\ \cline{2-3}
Novelty Seeking \footnote{adapted from Manning et al. \cite{Manning1995}} \newline(NOVS)  & novs\_2 & “I like to go to places where I will be exposed to information about new products and brands."   \\ \cline{2-3}
statements are equal & novs\_3 & “I like magazines that introduce new brands."   \\ \cline{2-3}
to and adopted from & novs\_4 & “I frequently look for new products and services."    \\ \cline{2-3}
Manning et al. \cite{Manning1995} & novs\_5 & “I seek out situations in which I will be exposed to new and different sources of product information."    \\ \cline{2-3}
& novs\_6 & “I am continually seeking new product experiences."     \\ \cline{2-3}
 & novs\_7 & “When I go shopping, I find myself spending very little time checking out new products and brands."     \\ \cline{2-3}
 & novs\_8 & “I take advantage of the first available opportunity to find out about new and different products."     \\ \hline
  & envc\_01 & “We are approaching the limit of the number of people the earth can support."     \\ \cline{2-3}
Environmental Concern\footnote{adapted from Dunlap et al. \cite{Dunlap2000} } \newline(ENVC) & envc\_02 & “Humans have the right to modify the natural environment to suit their needs."     \\ \cline{2-3}
statements are equal  & envc\_03 & “When humans interfere with nature it often produces disastrous consequences."    \\ \cline{2-3}
to and adopted from & envc\_04 & “Human ingenuity will insure that we do NOT make the earth unliveable."    \\ \cline{2-3}
Dunlap et al. \cite{Dunlap2000} & envc\_05 & “Humans are severely abusing the environment."     \\ \cline{2-3}
 & envc\_06 & “The earth has plenty of natural resources if we just learn how to develop them."     \\ \cline{2-3}
 & envc\_07 & “Plants and animals have as much right as humans to exist."       \\ \cline{2-3}
 & envc\_08 & “The balance of nature is strong enough to cope with the impacts of modern industrial nations."    \\ \cline{2-3}
 & envc\_09 & “Despite our special abilities humans are still subject to the laws of nature."   \\ \cline{2-3}
 & envc\_10 & “The so–called ‘‘ecological crisis’’ facing humankind has been greatly exaggerated."     \\ \cline{2-3}
 & envc\_11 & “The earth is like a spaceship with very limited room and resources."     \\ \cline{2-3}
 & envc\_12 & “Humans were meant to rule over the rest of nature."     \\ \cline{2-3}
 & envc\_13 & “The balance of nature is very delicate and easily upset."    \\ \cline{2-3}
 & envc\_14 & “Humans will eventually learn enough about how nature works to be able to control it."    \\ \cline{2-3}
 & envc\_15 & “If things continue on their present course, we will soon experience a major ecological catastrophe."  \\ \hline
 & att\_1 & I think I would feel good about adopting [a photovoltaic system $\mid$ an energy efficient fridge $\mid$ a green electricity tariff]   \\ \cline{2-3}
Attitude\footnote{based on the correspondence principle by \cite{Ajzen.1977}} \newline(ATT) & att\_2 & I think owning [a photovoltaic system $\mid$ an energy efficient fridge $\mid$ a green electricity tariff] is in line with my self-image.     \\ \cline{2-3}
 & att\_3 & I like the idea of having [a photovoltaic system $\mid$ an energy efficient fridge $\mid$ a green electricity tariff].    \\ \cline{2-3}
 & att\_4 & [Photovoltaic systems $\mid$ energy efficient fridges $\mid$ green electricity tariffs] are not for me.     \\ \hline
 & socn\_1 & For people in my situation in life, it is common to install [a photovoltaic system $\mid$ an energy efficient fridge $\mid$ a green electricity tariff].     \\ \cline{2-3}
Social Norm\footnote{modified from  \cite{KORCAJ2015407,schwarz2007umweltinnovationen}} \newline(SOCN) & socn\_2 & Many people in my circle of acquaintances would like it if I installed [a photovoltaic system $\mid$ an energy efficient fridge $\mid$ a green electricity tariff]. \\ \cline{2-3}
 & socn\_3 & Many people in my circle of acquaintances would like to buy/adopt [a photovoltaic system $\mid$ an energy efficient fridge $\mid$ a green electricity tariff] themselves if they could.      \\ \hline
 & finb\_1 & The costs associated with the adoption of [a photovoltaic system $\mid$ an energy efficient fridge $\mid$ a green electricity tariff] are too high for me.      \\ \cline{2-3}
Financial Benefit\footnote{modified from  \cite{schwarz2007umweltinnovationen,KORCAJ2015407,Wolske.2017}} \newline(FINB)  & finb\_2 & Overall, an investment in/ a conclusion of [a photovoltaic system $\mid$ an energy efficient fridge $\mid$ a green electricity tariff] is financially worthwhile for me.      \\ \cline{2-3}
 & finb\_3 & The investment in/ a conclusion of [a photovoltaic system $\mid$ an energy efficient fridge $\mid$ a green electricity tariff] pays for itself after a reasonable period of time.      \\ \cline{2-3}
 & finb\_4 & Instead of investing in/ paying for [a photovoltaic system $\mid$ an energy efficient fridge $\mid$ a green electricity], I would rather invest my money elsewhere.      \\ \hline
  & envb\_1 & I think that [photovoltaic systems $\mid$ energy efficient fridges $\mid$ green electricity tariffs] are a building block for a better environment.     \\ \cline{2-3}
Enviornmental Benefit\footnote{modified from  \cite{KORCAJ2015407,Mundaca2020,gerpott2010determinants}} \newline(ENVB)  & envb\_2 & The private use of [a photovoltaic system $\mid$ an energy efficient fridge $\mid$ a green electricity] would give me the feeling of doing something for the environment.   \\ \cline{2-3}
 & envb\_3 & In my opinion, other products are better suited to protecting the environment than [a photovoltaic system $\mid$ an energy efficient fridge $\mid$ a green electricity tariff].  \\ \hline
 & pbc\_1 & ...that my house deviates from the optimal conditions/ is equipped with another device / tariff \\ \cline{2-3}
Perceived Behavioural \footnote{based on the suggestions of \cite{ajzen2020}, and modified items of \cite{Wolske2017,Claudy.2013,Rai.2016}} \newline Control (PBC)   & pbc\_2 & ...that it is difficult to get good and comprehensive information about this technology / device / tariff.   \\ \cline{2-3}
 & pbc\_3 & ...that the time required for planning and installation seems too great to me. \\ \cline{2-3}
 & pbc\_4 &...that installation and operation are associated with too many risks and uncertainties.\\ \cline{2-3}
 & pbc\_5 &...that I do not like the optic of the technology / of the device.    \\ \cline{2-3}
& pbc\_6 &...that I don not believe that the technology is as environmentally friendly as it is advertised. \\ \cline{2-3}
 & pbc\_7 & ...that I have no idea how much longer I will be living in my current home.   \\ \cline{2-3}
 & pbc\_8 & ...that I think that the tariff change will no longer be worthwhile for me because of my age.    \\\cline{2-3}
 & pbc\_9 & ...that I can not afford green electricity.  \\ \cline{2-3}
 & pbc\_10 & ....that there are no devices/vendors/providers in my area.    \\ \hline
\end{longtable}
\normalsize

Finally, several personal questions were asked in order to enable comparison between different residential decision makers. The questions comprised socio-demographic questions (e.g., gender, education, income, household status, size,and composition); information about the area of residence (e.g., zip code, information about density and type of settlement, and urban or rural area); and details about the decision-making process (e.g., spark event, and decision period). Each of the questions contained prescribed response options in a single-choice format.

\subsection{Procedure of recruitment and sourced samples}
\label{S:3.2}

The sample for each \acrshort{lct} was sourced from the Respondi Online Access Panel using a specialised sampling tool that makes a random selection of potential participants. The quality control and testing were overseen by SINUS Markt- und Sozialforschung. While we checked the plausibility and consistency during the programming process, the questions and answer scales were further checked for comprehensibility during an external pre-test with 53 participants in May 2021. After revisions and modifications by us, the final data collection took place from 14 May to 31 May 2021 with the aid of three online surveys. For each of the three technologies – rooftop photovoltaic, energy-efficient fridge, and green electricity tariff – a separate sample of respondents was achieved who did not yet own the \acrshort{lct} in focus (\acrshort{pv}: N=1800; \acrshort{eea}: N=400; \acrshort{gt}: N=400). An overview of the socio-dempographic compositions is outlined in Table \ref{tab:sample}.

Each \acrshort{lct} sample was compared with structural data on German house owners using the best for planning tool \citep{SINUS_report.2021}. To determine how well each sample matched the population across different socio-demographic variables, a one-sample $\chi^{2}-$Test was conducted \citep{Bamberg2017}. 

It was demonstrated that that respondents in the \acrshort{pv}, \acrshort{eea}, and \acrshort{gt} sample are representative for the population of homeowners on the variables gender (\acrshort{pv}: $\chi^{2}=.04$; $p=.841$; \acrshort{eea}: $\chi^{2}=1.7$; $p=.193$; \acrshort{gt}: $\chi^{2}=.36$; $p=.548$), state (\acrshort{pv}: $\chi^{2}=8.36$; $p=.909$; \acrshort{eea}: $\chi^{2}=19.11$; $p=.209$; \acrshort{gt}: $\chi^{2}=24.46$; $p=.058$). Regarding the age group ($\chi^{2}=16.33$; $p=.096$), only the \acrshort{pv} sample was representative. 
At the same time, the conducted tests for the age group variable for the \acrshort{eea} and \acrshort{gt} sample, showed significantly differences to the population. Similarly, the variables family status, home ownership were significantly different for all \acrshort{lct} samples ($p\leq.001$). The inconsistencies in the variables might be explained by the survey design and operationalisation of the variables. Additionally, a descriptive comparison, as outlined in the Supplementary Material, shows that the sample differed only modestly.

\subsection{Model specifications and statistical analyses}
\label{S:3.3}

\acrlong{sem} (\acrshort{sem}) is a powerful technique widely used in social science research \citep{Ramlall2017}. Two approaches for the estimation of \acrshort{sem} can be distinguished: \acrlong{cbsem} (\acrshort{cbsem}) and \acrlong{plssem} (\acrshort{plssem}). We used \acrshort{plssem} to test the model due to its advantages over \acrshort{cbsem}, including the avoidance of distribution assumptions \citep{hair2017}, the estimation of composite scores \citep{Hair_primer_2017}, and its suitability for formative and single item measurement models \citep{Hair_primer_2017}. The path estimation for each of the three surveys (\acrshort{pv}, \acrshort{eea}, and \acrshort{gt}) and samples, as presented, was conducted using the package "plssem" by Venturini and Mehmetoglu \cite{venturini2019} for Stata 16. Following recommendation by Hair et al. \cite{Hair_primer_2017}, a maximum number of 100 iterations and seed number $seed=123$ were set \citep{Mehmetoglu}. 

After model respecifications, as detailed in the Supplementary Material, the global model provided sufficient reliability and validity estimates for measurement and structural models as described. The evaluation for the measurement and structural models is presented in Section \ref{S:4.1}. After calculating the results of the global sample of each \acrshort{lct} as outlined in \ref{S:4.2.1} and \ref{S:4.2.2}, we identified a similar predictor pattern across the high-, medium-, and low-cost \acrshort{lcts}, with particular emphasis on the relationships of the first and second hypotheses. The primary focus was on the perceived financial and environmental benefits.

In Section \ref{S:4.3}, we tested for differences in model parameters between socio-demographic groups in each \acrshort{lct} sample. In line with the hypothesis, we checked if household income, respondents age, or gender do not have a moderating effect on the relevant pathways underpinning attitudes to, and intention to adopt the three \acrshort{lcts} of interest. A non-parametric PLS-based approach to multi-group analysis was used \citep{venturini2019}. This approach first estimates the path coefficients and loadings for each group separately \citep{venturini2019} and then tests for significant differences between groups regarding the third hypothesis  \citep{Hair_primer_2017}. \par 

\par

\section{Results}
\label{S:4}

\subsection{Model evaluation}
\label{S:4.1}

For the reflective measurement models, the values of Cronbach's $\alpha$ and $\rho$ demonstrated high internal consistency reliability, with all constructs falling within the preferred range of 0.7 to 0.95 \citep{avkiran2018collec}. Concerning measures of convergent validity for the reflective constructs, all factor loadings were significant ($p<.001$) and surpassed the threshold of 0.5 \citep{Hair_primer_2017}. Discriminant validity was assessed using the Fornell-Larcker criterion \citep{Fornell.1981}, and all constructs met the necessary conditions, ensuring that each construct in the model is distinct and represents a specific latent concept.

Model evaluation for the measurement and the structural model are outlined in Table \ref{tab:evaluationresults}. For the reflective measurement models, the values of Cronbach's $\alpha$ and $\rho$ demonstrated high internal consistency reliability, with all constructs falling within the preferred range of 0.7 to 0.95 \citep{avkiran2018collec}. Concerning measures of convergent validity for the reflective constructs, all factor loadings were significant ($p<.001$) and surpassed the threshold of 0.5 \citep{Hair_primer_2017}. Discriminant validity was assessed using the Fornell-Larcker criterion \citep{Fornell.1981}, and all constructs met the necessary conditions, ensuring that each construct in the model is distinct and represents a specific latent concept.

In terms of formative measurement models and the structural model, there were no signs of collinearity, as confirmed by all \acrshort{vif} values being below the threshold of 5 \citep{Hair_primer_2017}. The predictive power was substantial for the endogenous construct ATT across all three models, with $R^{2}$ values ranging from 0.633 to 0.688. For the INT construct, the predictive power ranged from small ($R^{2} = 0.252$) to moderate ($R^{2} = 0.466-0.483$) \citep{Mehmetoglu}. 

Lastly, the overall quality of the structural model was satisfactory. The goodness-of-fit (\acrshort{gof}) was high across all models, with values of 0.969 for \acrshort{pv}, and 0.927 and 0.958 for \acrshort{eea} and \acrshort{gt}, respectively, indicating a satisfactory fit between the proposed model and the observed data.

\newpage
\scriptsize
\begin{table}[htp!]
\scriptsize
\centering
\caption{Evaluation criteria of the reflective and formative measurment models as well as the structural model for the applied \acrshort{sem} of \acrshort{pv}, \acrshort{eea}, and \acrshort{gt}. For a full display of the evaluation values and model estimates, see the Supplementary Material.}
\label{tab:evaluationresults}
\begin{tabular}{lllllllll}
                        \hline
\textbf{Criteria} & \textbf{ENVC} & \textbf{NOVS} & \textbf{FINB} & \textbf{ENVB} & \textbf{ATT} & \textbf{SOCN} & \textbf{PBC} & \textbf{INT} \\ \hline
\multicolumn{9}{c}{\textbf{PV}} \\ \hline      
alpha      & 0.883         & 0.942         & 0.692          & 0.778          & 0.923        & 0.789         &              &              \\
rho        & 0.901         & 0.944         & 0.743          & 0.841          & 0.924        & 0.792         &              &              \\
AVE       & 0.46          & 0.745         & 0.521          & 0.7            & 0.812        & 0.704         &              &              \\
VIF      &               &               &                &                &              &               & 1.07 - 2.07    &              \\
R2      &               &               &                &                & 0.633        &               &              & 0.466        \\
VIF    & 1.18          & 1.094         & 1.465          & 1.617          &              &               &              &              \\
    &               &               & 1.796          & 2.229          & 3.336        & 2.059         & 1.282        &              \\
GoF    &               &               &                &                &              &               &              & 0.969        \\ \hline
\multicolumn{9}{c}{\textbf{EEA}}                                                                                                                 \\ \hline
alpha     & 0.889         & 0.945         & 0.786          & 0.681          & 0.844        & 0.814         &              &              \\
rho      & 0.911         & 0.95          & 0.809          & 0.852          & 0.902        & 0.855         &              &              \\
AVE       & 0.452         & 0.752         & 0.606          & 0.632          & 0.697        & 0.73          &              &              \\
VIF        &               &               &                &                &              &               & 1.07 - 3.65    &              \\
R2       &               &               &                &                & 0.664        &               &              & 0.252        \\
VIF      & 1.375         & 1.16          & 1.589          & 1.91           &              &               &              &              \\
    &               &               & 1.94           & 2.538          & 3.271        & 2.022         & 1.083        &              \\
GoF     &               &               &                &                &              &               &              & 0.927        \\ \hline
\multicolumn{9}{c}{\textbf{GT}}                                                                                                                  \\ \hline
alpha     & 0.897         & 0.943         & 0.679          & 0.822          & 0.911        & 0.818         &              &              \\
rho        & 0.927         & 0.96          & 0.694          & 0.891          & 0.913        & 0.821         &              &              \\
AVE        & 0.471         & 0.746         & 0.506          & 0.741          & 0.79         & 0.733         &              &              \\
VIF        &               &               &                &                &              &               & 1.24 - 2.54    &              \\
R2     &               &               &                &                & 0.68         &               &              & 0.483        \\
VIF      & 1.361         & 1.041         & 1.371          & 1.715          &              &               &              &              \\
    &               &               & 1.613          & 3.007          & 3.495        & 1.935         & 1.434        &              \\
GoF      &               &               &                &                &              &               &              & 0.958       
\end{tabular}
\end{table}
\normalsize

\subsection{Model results}

Based on the positive model evaluation described above, Figure \ref{fig:PathCoefficients} presents the path diagram illustrating the model variance and estimated coefficients for each LCT. For an exhaustive view of the model outcomes, readers are directed to the Supplementary Material. Additionally, the interrelations between the latent variables of the augmented \acrshort{tpb} are tabulated in Tables \ref{tab:CorrPV}, \ref{tab:CorrEEA}, and \ref{tab:CorrGT}.

\subsubsection{Attitudes toward low-carbon technologies}
\label{S:4.2.1}
\begin{figure}[htp!]
    \centering
    \subfloat[\acrshort{pv}]{\includegraphics[width=0.85\textwidth]{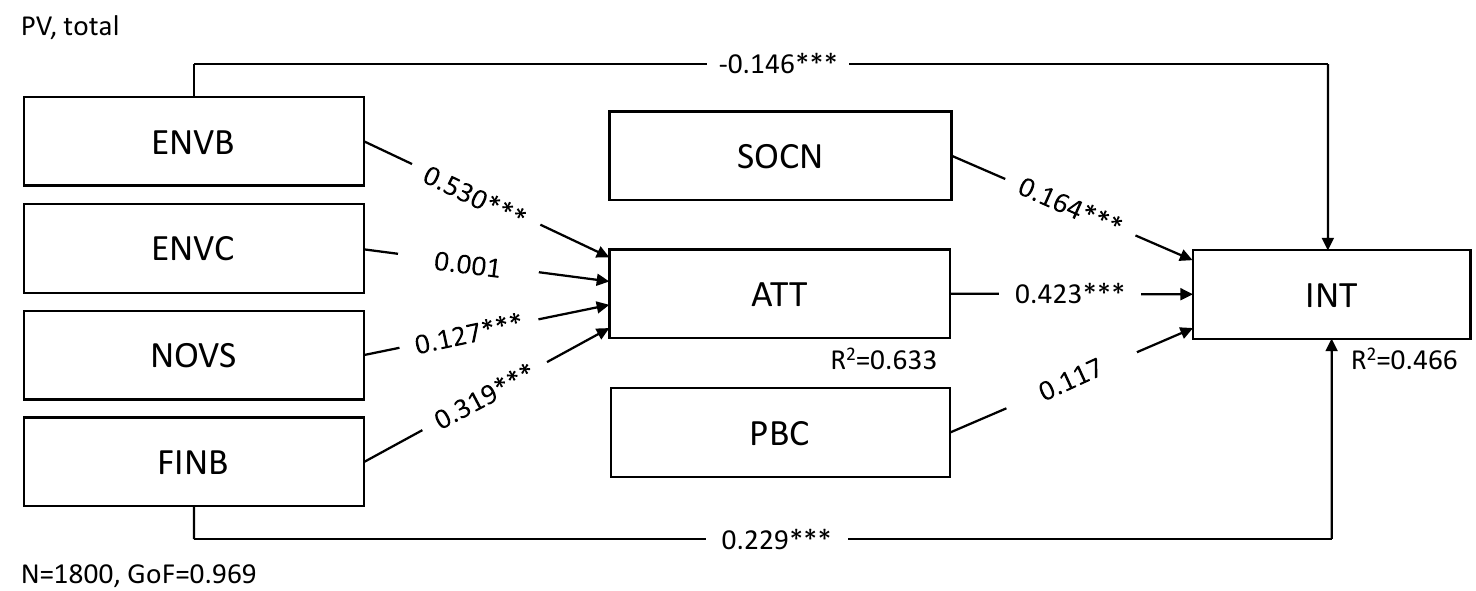}}\\
    \subfloat[\acrshort{eea}]{\includegraphics[width=0.85\textwidth]{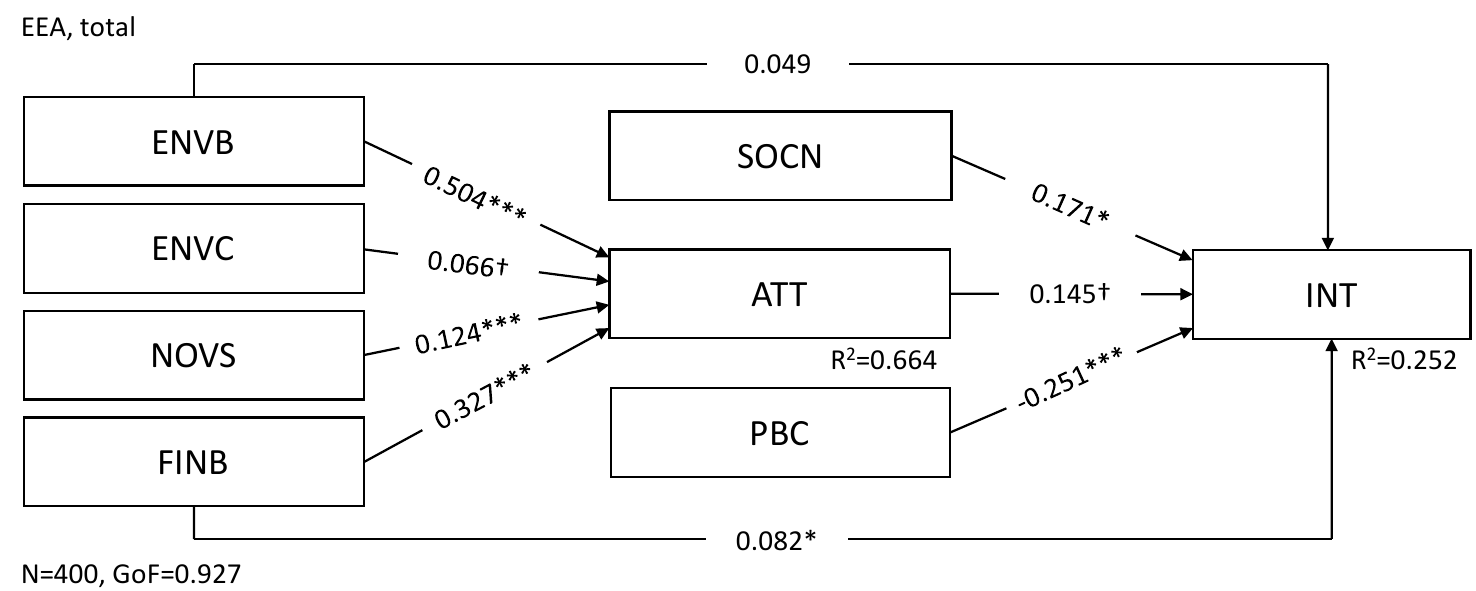}}
    \\
    \subfloat[\acrshort{gt}]{\includegraphics[width=0.85\textwidth]{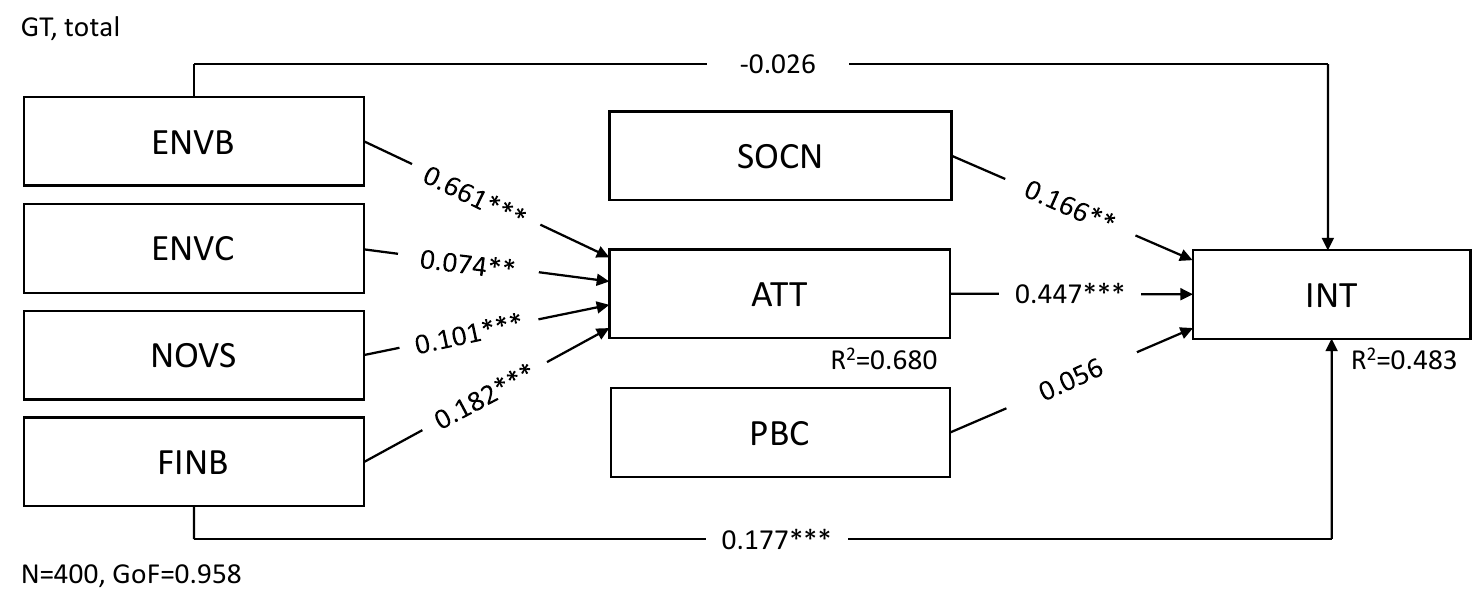}}
    \caption{Estimated path coefficients of the \acrshort{tpb} for the total sample of each of the three assessed \acrshort{lcts} (\acrshort{pv}, \acrshort{eea}, \acrshort{gt}). Statistical significance is indicated using crosses and asterisks (\textsuperscript{$\dagger$}$p<.100$, \textsuperscript{*}$p<.050$, \textsuperscript{**}$p<.010$, \textsuperscript{***}$p<.001$). While environmental benefits (ENVB) was a better predictor for explaining a positive behavioural attitude towards adopting the respective technology than financial benefits, financial benefits (FINB) had a higher coefficient than environmental benefits for explaining a positive intention to adopt the respective technology. This pattern was visible for each of the \acrshort{lct}}.
    \label{fig:PathCoefficients}
\end{figure}

Demonstrating the strong predictive power of our sele constructs, over two-thirds of the variance in ATT ($R^{2}_{PV}$\,=\,0.663, $R^{2}_{EEA}$\,=\,0.664, $R^{2}_{GT}$\,=\,0.681) was explained by the exogenous constructs NOVS, FINB, ENVB, and ENVC (Figure \ref{fig:PathCoefficients}). Notably, both benefit constructs, ENVB and FINB were found to be strong predictors of ATT for all three \acrshort{lcts}. As hypothesised ENVB (H1) ($\beta_{PV}\,=\,0.530$,$\beta_{EEA}\,=\,0.504$,$\beta_{GT}\,=\,0.661$) demonstrated a higher predictive value than FINB for all \acrshort{lcts} ($\beta_{PV}\,=\,0.319$,$\beta_{EEA}\,=\,0.327$,$\beta_{GT}\,=\,0.183$). 

\subsubsection{Intention to adopt low-carbon technologies}
\label{S:4.2.2}

Concerning the intention to adopt, the exogenous constructs SOCN, ATT, PBC, FINB, and ENVB demonstrated substantial predictive power for \acrshort{pv} and \acrshort{gt} ($R^{2}_{PV}$\,=\,0.466, $R^{2}_{GT}$\,=\,0.483), accounting for roughly 50\,\% of the variance in INT to adopt the 2 technologies. However, for \acrshort{eea}, a significantly lower variance was observed ($R^{2}_{eea}$\,=\,0.252). The adoption INT of the three selected \acrshort{lcts} was generally influenced by similar determinants. The relationships between SOCN and ATT on INT were positive and significant for \acrshort{pv} and \acrshort{gt}, with a similar relationship noted for \acrshort{eea} at the 10\% level. Differences in the sign, strength, and significance of the relationship coefficient between PBC and INT varied amongst the three \acrshort{lcts}. 

Regarding the two benefit constructs, FINB played a pivotal role in explaining variance in INT to adopt across each of the \acrshort{lcts} ($\beta_{PV}\,=\,0.229$,$\beta_{EEA}\,=\,0.082$,$\beta_{GT}\,=\,0.191$). Conversely, the coefficients for ENVB were not significant for \acrshort{eea} and \acrshort{gt} ($\beta_{EEA}\,=\,0.049$,$\beta_{GT}\,=\,-0.191$), whereas the relationship between adoption INT and ENVB for \acrshort{pv} ($\beta_{PV}\,=\,-0.146$,) was negative. These results corroborate our hypothesis (H2) that as homeowners progress through the decision-making process concerning the adoption of \acrshort{lcts} according to the \acrshort{tpb}, perceived financial benefits become a more salient predictor of intention to adopt than perceived environmental benefits.

\subsubsection{Sub-group analysis}
\label{S:4.3}

To scrutinise our hypothesis across different population segments, Table \ref{tab:SubgroupRelationships} provides the relevant path coefficients for the relationships between ENVB and FINB and attitude, and ENVB and FINB and intention to adopt for each of the three \acrshort{lcts} categorised by income, age, and gender. Multi-group comparisons between the sub-groups of one \acrshort{lct} were also conducted to test for significant variations between the groups (H3). The complete estimated relationships between all constructs for each of the three models can be found in the Supplementary Material. 

Table \ref{tab:SubgroupRelationships} demonstrates that the estimated path coefficients for the two homeowner income brackets (high and low) and by gender (men and women) largely reflect the population level outcomes for all three \acrshort{lcts}. In line with the primary findings, ENVB has a more potent influence on ATT, whilst FINB exerts a greater influence on the intention to adopt (INT). Pairwise comparisons did not disclose any notable differences between the sub-groups at the 5\% significance level. 

Concerning the sub-group analysis by age, we again observed that the estimated path coefficients for each of the three age brackets largely mirror the overall population model. However, the estimated coefficients for \acrshort{eea} are notably divergent from one another. Specifically, the coefficients for FINB on ATT are considerably weaker for respondents aged 18-39Y and 40-59Y in comparison to the eldest cohort (60-79Y) ($p=0.023$, $p=0.036$, respectively). Moreover, the impact of FINB on INT is significantly reduced for respondents aged 18-39Y compared to those aged 60-79Y ($p=0.021$).

In summary, these findings support our hypothesis that socio-demographic attributes do not moderate the pertinent pathways underpinning attitudes towards, and intentions to adopt the three \acrshort{lcts} in question (H3). Irrespective of household income, age, and gender, perceived financial benefits rise as a more influential predictor of intention to adopt than perceived environmental benefits as homeowners navigate the decision-making process concerning \acrshort{lct} adoption.

\newpage
\scriptsize
% \usepackage{multirow}
% \usepackage{longtable}
% Note: It may be necessary to compile the document several times to get a multi-page table to line up properly
\begin{longtable}[c]{llllcllc}
\caption{Estimated path coefficients for the relationships between ENVB or FINB and ATT, and for the relationships between ENEV or FINB and INT for the total sample and different sub-samples in the \acrshort{pv}, \acrshort{eea}, and \acrshort{gt} surveys. The p-values for statistical significance are g in parentheses. When the direction of hypotheses and the significance level are confirmed, there is a strong support of our hypothesis (++). When only the direction is fulfilled, there is a weak support (+). The negative signes are vice versa.}\\ 
\label{tab:SubgroupRelationships}\\
\hline
\multirow{2}{*}{\textbf{Sample}} & \multicolumn{7}{c}{\textbf{Path-relationship}}                                                                                                                                                                                                                                                       \\ \cline{2-8} 
                                 & \textbf{N} & \multicolumn{1}{c}{\textbf{ENVB $\rightarrow$ ATT}}        & \multicolumn{1}{c}{\textbf{FINB $\rightarrow$  ATT}}        & \textbf{H1} & \multicolumn{1}{c}{\textbf{ENVB $\rightarrow$  INT}}         & \multicolumn{1}{c}{\textbf{FINB $\rightarrow$ INT}}         & \textbf{H2} \\ \hline
\endfirsthead
\endhead
\hline
\endfoot
\endlastfoot
\multicolumn{8}{c}{PV}                                                                                                                                                                                                                                                                                                                  \\ \hline
Total                            & 1.800      & \begin{tabular}[c]{@{}l@{}}0.530\\      (0.000)\end{tabular} & \begin{tabular}[c]{@{}l@{}}0.319\\      (0.000)\end{tabular} & ++          & \begin{tabular}[c]{@{}l@{}}-0.146\\      (0.000)\end{tabular} & \begin{tabular}[c]{@{}l@{}}0.229 \\      (0.000)\end{tabular} & ++          \\
\multicolumn{8}{l}{Income}                                                                                                                                                                                                                                                                                                              \\ \hline
LowInc                           & 989        & \begin{tabular}[c]{@{}l@{}}0.506\\      (0.000)\end{tabular} & \begin{tabular}[c]{@{}l@{}}0.346\\      (0.000)\end{tabular} & ++          & \begin{tabular}[c]{@{}l@{}}-0.170\\      (0.000)\end{tabular} & \begin{tabular}[c]{@{}l@{}}0.215\\      (0.000)\end{tabular}  & ++          \\
HighInc                          & 811        & \begin{tabular}[c]{@{}l@{}}0.554\\      (0.000)\end{tabular} & \begin{tabular}[c]{@{}l@{}}0.292\\      (0.000)\end{tabular} & ++          & \begin{tabular}[c]{@{}l@{}}-0.111\\      (0.003)\end{tabular} & \begin{tabular}[c]{@{}l@{}}0.243\\      (0.000)\end{tabular}  & ++          \\
\multicolumn{8}{l}{Age}                                                                                                                                                                                                                                                                                                                 \\ \hline
18-39Y                           & 479        & \begin{tabular}[c]{@{}l@{}}0.554\\      (0.000)\end{tabular} & \begin{tabular}[c]{@{}l@{}}0.262\\      (0.000)\end{tabular} & ++          & \begin{tabular}[c]{@{}l@{}}-0.095\\      (0.049)\end{tabular} & \begin{tabular}[c]{@{}l@{}}0.244\\      (0.000)\end{tabular}  & ++          \\
40-59Y                           & 804        & \begin{tabular}[c]{@{}l@{}}0.540\\      (0.000)\end{tabular} & \begin{tabular}[c]{@{}l@{}}0.336\\      (0.000)\end{tabular} & ++          & \begin{tabular}[c]{@{}l@{}}-0.149\\      (0.000)\end{tabular} & \begin{tabular}[c]{@{}l@{}}0.272\\      (0.000)\end{tabular}  & ++          \\
60-79Y                           & 517        & \begin{tabular}[c]{@{}l@{}}0.532\\      (0.000)\end{tabular} & \begin{tabular}[c]{@{}l@{}}0.328\\      (0.000)\end{tabular} & ++          & \begin{tabular}[c]{@{}l@{}}-0.150\\      (0.000)\end{tabular} & \begin{tabular}[c]{@{}l@{}}0.155\\      (0.000)\end{tabular}  & ++          \\
\multicolumn{8}{l}{Gender}                                                                                                                                                                                                                                                                                                              \\ \hline
Female                           & 950        & \begin{tabular}[c]{@{}l@{}}0.525\\      (0.000)\end{tabular} & \begin{tabular}[c]{@{}l@{}}0.330\\      (0.000)\end{tabular} & ++          & \begin{tabular}[c]{@{}l@{}}-0.101\\      (0.004)\end{tabular} & \begin{tabular}[c]{@{}l@{}}0.220\\      (0.000)\end{tabular}  & ++          \\
Male                             & 850        & \begin{tabular}[c]{@{}l@{}}0.538\\      (0.000)\end{tabular} & \begin{tabular}[c]{@{}l@{}}0.308\\      (0.000)\end{tabular} & ++          & \begin{tabular}[c]{@{}l@{}}-0.184\\      (0.000)\end{tabular} & \begin{tabular}[c]{@{}l@{}}0.245\\      (0.000)\end{tabular}  & ++         
                                                                                                                                      \\ \hline
\multicolumn{8}{c}{EEA}                                                                                                                                                                                                                                                                                                                 \\ \hline
Total                            & 400        & \begin{tabular}[c]{@{}l@{}}0.504\\      (0.000)\end{tabular} & \begin{tabular}[c]{@{}l@{}}0.327\\      (0.000)\end{tabular} & ++          & \begin{tabular}[c]{@{}l@{}}0.049\\      (0.510)\end{tabular}  & \begin{tabular}[c]{@{}l@{}}0.082\\      (0.246)\end{tabular}  & +           \\
\multicolumn{8}{l}{Income}                                                                                                                                                                                                                                                                                                              \\ \hline
LowInc                           & 222        & \begin{tabular}[c]{@{}l@{}}0.479\\      (0.000)\end{tabular} & \begin{tabular}[c]{@{}l@{}}0.336\\      (0.000)\end{tabular} & ++          & \begin{tabular}[c]{@{}l@{}}0.021\\      (0.824)\end{tabular}  & \begin{tabular}[c]{@{}l@{}}0.146\\      (0.118)\end{tabular}  & +           \\
HighInc                          & 178        & \begin{tabular}[c]{@{}l@{}}0.539\\      (0.000)\end{tabular} & \begin{tabular}[c]{@{}l@{}}0.320\\      (0.000)\end{tabular} & ++          & \begin{tabular}[c]{@{}l@{}}0.086\\      (0.459)\end{tabular}  & \begin{tabular}[c]{@{}l@{}}0.016\\      (0.863)\end{tabular}  & -           \\
\multicolumn{8}{l}{Age}                                                                                                                                                                                                                                                                                                                 \\ \hline
18-39Y                           & 131        & \begin{tabular}[c]{@{}l@{}}0.565\\      (0.000)\end{tabular} & \begin{tabular}[c]{@{}l@{}}0.277\\      (0.000)\end{tabular} & ++          & \begin{tabular}[c]{@{}l@{}}-0.008\\      (0.951)\end{tabular} & \begin{tabular}[c]{@{}l@{}}-0.132\\      (0.312)\end{tabular} & -           \\
40-59Y                           & 164        & \begin{tabular}[c]{@{}l@{}}0.532\\      (0.000)\end{tabular} & \begin{tabular}[c]{@{}l@{}}0.237\\      (0.000)\end{tabular} & ++          & \begin{tabular}[c]{@{}l@{}}0.066\\      (0.560)\end{tabular}  & \begin{tabular}[c]{@{}l@{}}0.115\\      (0.249)\end{tabular}  & +           \\
60-79Y                           & 105        & \begin{tabular}[c]{@{}l@{}}0.364\\      (0.004)\end{tabular} & \begin{tabular}[c]{@{}l@{}}0.530\\      (0.000)\end{tabular} & --          & \begin{tabular}[c]{@{}l@{}}-0.113\\      (0.431)\end{tabular} & \begin{tabular}[c]{@{}l@{}}0.209\\      (0.139)\end{tabular}  & +           \\
\multicolumn{8}{l}{Gender}                                                                                                                                                                                                                                                                                                              \\ \hline
Female                           & 225        & \begin{tabular}[c]{@{}l@{}}0.523\\      (0.000)\end{tabular} & \begin{tabular}[c]{@{}l@{}}0.272\\      (0.000)\end{tabular} & ++          & \begin{tabular}[c]{@{}l@{}}0.119\\      (0.225)\end{tabular}  & \begin{tabular}[c]{@{}l@{}}0.032\\      (0.732)\end{tabular}  & -           \\
Male                             & 175        & \begin{tabular}[c]{@{}l@{}}0.471\\      (0.000)\end{tabular} & \begin{tabular}[c]{@{}l@{}}0.379\\      (0.000)\end{tabular} & ++          & \begin{tabular}[c]{@{}l@{}}-0.044\\      (0.672)\end{tabular} & \begin{tabular}[c]{@{}l@{}}0.225\\      (0.017)\end{tabular}  & ++          \\ \hline
\multicolumn{8}{c}{GT}                                                                                                                                                                                                                                                                                                                  \\ \hline
Total                            & 400        & \begin{tabular}[c]{@{}l@{}}0.661\\      (0.000)\end{tabular} & \begin{tabular}[c]{@{}l@{}}0.182\\      (0.000)\end{tabular} & ++          & \begin{tabular}[c]{@{}l@{}}-0.026\\      (0.695)\end{tabular} & \begin{tabular}[c]{@{}l@{}}0.177\\      (0.000)\end{tabular}  & ++          \\
\multicolumn{8}{l}{Income}                                                                                                                                                                                                                                                                                                              \\ \hline
LowInc                           & 223        & \begin{tabular}[c]{@{}l@{}}0.639\\      (0.000)\end{tabular} & \begin{tabular}[c]{@{}l@{}}0.150\\      (0.021)\end{tabular} & ++          & \begin{tabular}[c]{@{}l@{}}-0.043\\      (0.604)\end{tabular} & \begin{tabular}[c]{@{}l@{}}0.216\\      (0.001)\end{tabular}  & ++          \\
HighInc                          & 177        & \begin{tabular}[c]{@{}l@{}}0.687\\      (0.000)\end{tabular} & \begin{tabular}[c]{@{}l@{}}0.234\\      (0.000)\end{tabular} & ++          & \begin{tabular}[c]{@{}l@{}}0.005\\      (0.961)\end{tabular}  & \begin{tabular}[c]{@{}l@{}}0.134\\      (0.068)\end{tabular}  & +           \\
\multicolumn{8}{l}{Age}                                                                                                                                                                                                                                                                                                                 \\ \hline
18-39Y                           & 129        & \begin{tabular}[c]{@{}l@{}}0.659\\      (0.000)\end{tabular} & \begin{tabular}[c]{@{}l@{}}0.206\\      (0.001)\end{tabular} & ++          & \begin{tabular}[c]{@{}l@{}}0.060\\      (0.628)\end{tabular}  & \begin{tabular}[c]{@{}l@{}}0.192\\      (0.002)\end{tabular}  & ++          \\
40-59Y                           & 155        & \begin{tabular}[c]{@{}l@{}}0.635\\      (0.000)\end{tabular} & \begin{tabular}[c]{@{}l@{}}0.210\\      (0.002)\end{tabular} & ++          & \begin{tabular}[c]{@{}l@{}}-0.216\\      (0.033)\end{tabular} & \begin{tabular}[c]{@{}l@{}}0.110\\      (0.207)\end{tabular}  & +           \\
60-79Y                           & 105        & \begin{tabular}[c]{@{}l@{}}0.683\\      (0.000)\end{tabular} & \begin{tabular}[c]{@{}l@{}}0.128\\      (0.142)\end{tabular} & ++          & \begin{tabular}[c]{@{}l@{}}0.026\\      (0.830)\end{tabular}  & \begin{tabular}[c]{@{}l@{}}0.244\\      (0.003)\end{tabular}  & ++          \\
\multicolumn{8}{l}{Gender}                                                                                                                                                                                                                                                                                                              \\ \hline
Female                           & 206        & \begin{tabular}[c]{@{}l@{}}0.688\\      (0.000)\end{tabular} & \begin{tabular}[c]{@{}l@{}}0.222\\      (0.000)\end{tabular} & ++          & \begin{tabular}[c]{@{}l@{}}0.030\\      (0.761)\end{tabular}  & \begin{tabular}[c]{@{}l@{}}0.164\\      (0.011)\end{tabular}  & ++          \\
Male                             & 194        & \begin{tabular}[c]{@{}l@{}}0.617\\      (0.000)\end{tabular} & \begin{tabular}[c]{@{}l@{}}0.153\\      (0.026)\end{tabular} & ++          & \begin{tabular}[c]{@{}l@{}}-0.090\\      (0.300)\end{tabular} & \begin{tabular}[c]{@{}l@{}}0.178\\      (0.011)\end{tabular}  & ++          \\ \hline
\end{longtable}
\normalsize

\section{Discussion}
\label{S:5}

\subsection{Principal findings}
Understanding the attitudes towards \acrshort{lcts} and the intention to adopt them among private homeowners is crucial for the formulation of evidence-based policies. Employing a similar survey instrument to assess the relative importance of collective versus homeowner benefits throughout the decision-making process for three residential \acrshort{lcts} — \acrshort{pv}, \acrshort{eea}, and \acrshort{gt} — our findings suggest that, irrespective of the cost base (whether non-cost products or high-cost products), homeowner attitudes and intentions to adopt these technologies are generally informed by similar determinants. Notably, product-specific benefits, as opposed to affective constructs, demonstrated a more robust relationship with attitudes towards \acrshort{lcts}, based on our chosen method of \acrshort{plssem}. Recognising that a positive attitude and intention are precursors to the eventual adoption of \acrshort{lcts}, it becomes evident that personal benefits and an emphasis on self-interest have a more significant impact than collective benefits.

Examining product-specific benefits, the adoption of \acrshort{lcts} by homeowners offers both collective benefits to society through reduced carbon emissions and individual financial benefits through lower energy costs. Using the \acrshort{tpb} and \acrshort{plssem}, we found that the perceived environmental (collective) benefits associated with the three \acrshort{lcts} had a stronger relationship at the attitude formation stage compared to perceived financial benefits. However, as homeowners progress through the decision-making process, the perceived financial benefits of the technology emerge as a more salient evaluative criterion. While both perceived financial and environmental benefits significantly influenced attitude formation for each of the three \acrshort{lcts}, the environmental benefits appeared to have a negligible association at the intention formation stage.

Although product-specific benefits played a greater role in attitude and intention formation for our three selected \acrshort{lcts}, interesting findings regarding the affective constructs and barriers were also found. Here we found that novelty seeking is a significant predictor for attitude formation, this relationship was weak for all three \acrshort{lcts}. Regarding \acrshort{pbc}, differences for each \acrshort{lct} arose, with barriers related to \acrshort{pv} and \acrshort{gt} having an insignificant moderate and weak association with adoption intention. For \acrshort{eea}, the perceived behavioural control construct was not seen as an actual barrier taking the other constructs into account. Regarding the affective constructs, the direction and significance of the coefficients are in line with previous research \cite{Klockner.2013,Bamberg2017,Claudy.2013}; however, our coefficients are slightly lower. This is surprising given that \acrshort{pv} systems in particular are subject to a large number of barriers compared to other pro-environmental behaviours such as recycling, water-saving or meat consumption. 

Overall, these results support the hypothesis that whilst environmental attitudes are important in developing a positive homeowners' attitude towards \acrshort{lcts} (H1), as homeowners progress through the decision-making process regarding the adoption of \acrshort{lcts}, the perceived financial benefits associated with the technology become a more crucial evaluative criterion (H2). Regarding H3, multi-group \acrshort{plssem} demonstrated that the relationships underpinning attitude formation and intention formation were generally consistent across different socio-demographic and economic sub-groups (16 out of 18 sub-group analyses), including income, age, and gender. Despite minor fluctuations in the magnitude of the pathway coefficients during the sub-group evaluation, the trajectory and significance of the constructs remained largely consistent across the different \acrshort{lcts}. This consistency underscores existing research, which indicates that gender, income, and age do not markedly influence attitudes or intentions to adopt \acrshort{lcts} \citep{Kastner.2015,Klockner.2013,Arts.2011}.

\subsection{Policy implications}
From a policy perspective, this study found that irrespective of initial cost, the environmental and financial product-specific benefits are an important determinant of homeowner's attitudes and intentions to adopt \acrshort{lcts}. Therefore, practical and policy measures should prioritize enhancing the perception of these benefits. It is important to note that,  as homeowners progress through the adoption decision-making process, the perceived individual financial returns to the house owners of the \acrshort{lct} were a more important determinate than the collective, environmental benefits as homeowners proceeded through the adoption decision-making process. However, strong perceived environmental benefits were key to homeowners having a positive attitude to the \acrshort{lct} in the first place and encouraging homeowner adoption. Policymakers should continue to emphasize the environmental benefits of \acrshort{lcts}, highlighting their contribution to carbon emissions reduction and overall sustainability. Communicating the broader societal and environmental impacts of \acrshort{lct} adoption can reinforce homeowners' positive attitudes and intentions. 

To promote \acrshort{lcts} effectively, population-based strategies should focus on both the collective (environmental) and individual (financial) level benefits they offer. A potential approach to achieve this could involve developing a series of 'representative' or archetypal house owners \cite{williams2021fostering}, demonstrating the financial savings they could expect based on their specific homeowner characteristics upon adopting an \acrshort{lcts}. Additionally, government-funded intermediaries, such as Smart Energy GB in the United Kingdom or Energiesprong in the Netherlands, have proven effective in connecting homeowners and businesses to the necessary skills and knowledge for transitioning to \acrshort{lcts} and practices \cite{sovacool2020guides}. Policymakers should consider establishing or supporting such intermediaries to facilitate the adoption process. These intermediaries can provide guidance, resources, and access to financing options, making it easier for homeowners to adopt \acrshort{lcts}.

Furthermore, to increase the financial benefits associated with \acrshort{pv}, \acrshort{eea}, and \acrshort{gt}, measures should not only focus on reducing initial costs but also provide clear explanations of the potential savings in energy prices and the amortisation period relative to other technology options. Clear and accurate information can help homeowners make informed decisions and understand the long-term financial advantages of adopting \acrshort{lcts}.

While product-specific benefits play a significant role in attitude and intention formation, it is crucial to address barriers that hinder the adoption of \acrshort{lcts}. For example, barriers should be identified and addressed through supportive policies, financial incentives, and simplified installation processes. Additionally, promoting novelty seeking can be an effective strategy to enhance homeowners' positive attitudes toward \acrshort{lcts}. Innovative marketing campaigns, information dissemination, and showcasing success stories can help overcome resistance to change and encourage adoption.

\subsection{Limitations}
Inherent to survey-based research is the potential for respondents to evaluate their perceptions and intentions differently depending on the time and context \citep{Wolske.2020}. In this study's context, German respondents were invited to discuss the perceived benefits, attitudes, and intentions regarding one of the \acrshort{lcts}. Consequently, it remains uncertain at which stage, if at all, our sampled homeowners are in the decision-making process even though actual adopters have been excluded. Moreover, the decision to exclude respondents unaware of the topic has constrained the spectrum of comparative analysis that could be undertaken and reduced the ability to account for self-selection bias. In terms of our sample's representativeness, it was sourced from the respondi Online Access Panel utilising a specialised sampling tool (EFS) for a random participant selection. Given the institute employs a consistent panel, the potential for self-selection bias remains.

Though we employed an expanded \acrshort{tpb} as the foundation for all three sub-samples to maintain comparability, slight disparities in the questions of the three survey instruments emerged due to the inherent characteristics of the \acrshort{lcts}, which affected their comparability. Minor specifications were also required for each \acrshort{lct} model in relation to the question. From a statistical perspective, we relied on the Fornell-Larcker criterion as evaluation criteria and not on HTMT, which is seen as more appropriate for examining discriminant validity \citep{benitez2020}. The reason was that the Stata package did not support the other tests.

In conclusion, our research did not delve into every conceivable influence on attitude or intention. To prevent overwhelming respondents, only one personal and one collective benefit were incorporated into the survey, in contrast to studies like \cite{KORCAJ2015407}. The intention variable was evaluated using single-item measures that may be prone to measurement error and might thus have impacted our conclusions. Despite of the limitations, our analyses demonstrate similar results across all three \acrshort{lcts}. Yet, notwithstanding these constraints, the attained $R^{2}$ values for the internal variables surpassed the average from past studies ($R^{2}$,=,0.39) utilising the \acrshort{tpb}, as highlighted in a meta-analysis by Armitage and Conner \cite{Armitage.2001}.

\section{Conclusion}
\label{S:6}

\acrshort{lcts} encompass a broad spectrum of technologies, ranging from no-cost services like \acrshort{gt}, to medium-cost offerings such as \acrshort{eea}, and premium products like \acrshort{pv}. Diverging from prior studies that narrow their lens to singular \acrshort{lcts}, our research provides a panoramic view. We unveil that despite different cost spectrums, the adoption intentions for \acrshort{pv}, \acrshort{gt} and \acrshort{eea} were generally underpinned by similar determinants. We also demonstrated that socio-demographic characteristics do not have a moderating effect on the relevant pathways.Finally, our findings have revealed a significant and positive correlation between the perception of financial benefits and one's attitude or intention to adopt \acrshort{pv}, \acrshort{eea}, and \acrshort{gt}. To simplify, if individuals hold the belief that these technologies can lead to cost savings over time, their inclination to embrace them becomes stronger. This observation highlights a prevailing tendency towards motivations rooted in self-interest, often characterized as greedy behavior, as well as concerns centered around personal financial gain rather than giving precedence to environmentally conscious choices, often referred to as green behavior. This underlines the necessity for an enhanced emphasis on promoting awareness regarding the accompanying individual or future financial advantages inherent in \acrshort{lcts}. As a result, there arises a clear need to inform individuals about the potential for reduced long-term expenses through the utilization of such technologies.

Overall, this study made an important contribution to previous research in this area. Equating outcomes across diverse studies remains an intricate endeavor, given the heterogeneity in sampling modalities and the conceptualization of constructs. By using the same extended \acrshort{tpb} with similarly operationalised constructs, we were able to identify similar patterns in explaining the intention to adopt across low- and high-cost \acrshort{lcts}, especially when looking at the associations with the perceived personal or collective benefits and attitudes and intention. Subsequent research could validate our assertions across varied \acrshort{lcts}. Delving into adoption surveys centered around energy-efficient insulations, e-vehicles, energy-conserving bulbs, or energy management frameworks might further elucidate the universal applicability of our conclusions. A propitious avenue for upcoming studies would be to scrutinize the relationship between perceived fiscal and ecological advantages, as pinpointed in our study, across the entirety of the decision-making continuum.

\section*{Acknowledgement}
Fabian Scheller kindly acknowledges the financial support of the European Union's Horizon 2020 research and innovation programme under the Marie Sklodowska-Curie grant agreement no. 713683 (COFUNDfellowsDTU). Furthermore, parts of the surveys were sponsored by the SUSIC project (Smart Utilities and Sustainable Infrastructure Change) with the project number 100378087. The project is financed by the Saxon State government out of the State budget approved by the Saxon State Parliament.

\bibliography{ModelDescription.bib}
%Bibstyle: numerical, first citation=1
\bibliographystyle{elsarticle-num}
%Bibstyle: alphabetical by last name
%\bibliographystyle{abbrvnat}

%% The Appendices part is started with the command \appendix; appendix sections are then done as normal sections
\appendix
\newpage
\section{Samples}

\scriptsize
% \usepackage{longtable}
% Note: It may be necessary to compile the document several times to get a multi-page table to line up properly
\begin{longtable}[c]{lrrrrrr}
\caption{Sample composition of \acrshort{pv}, \acrshort{eea}, and \acrshort{pv} survey}
\label{tab:sample}\\
\hline
Socio-demographic &
  \multicolumn{2}{l}{PV} &
  \multicolumn{2}{l}{EEA} &
  \multicolumn{2}{l}{GT} \\
\endfirsthead
\endhead
variables &
  \multicolumn{1}{l}{N} &
  \multicolumn{1}{l}{\%} &
  \multicolumn{1}{l}{N} &
  \multicolumn{1}{l}{\%} &
  \multicolumn{1}{l}{N} &
  \multicolumn{1}{l}{\%} \\ \hline
\multicolumn{7}{c}{Gender}                                                \\ \hline
Female                   & 950  & 52.78\% & 225 & 56.25\% & 206 & 51.50\% \\
Male                     & 850  & 47.22\% & 175 & 43.75\% & 194 & 48.50\% \\ \hline
\multicolumn{7}{c}{Age group}                                             \\ \hline
18 - 29 Years            & 199  & 11.06\% & 78  & 19.50\% & 77  & 19.25\% \\
30 - 39 Years            & 280  & 15.56\% & 53  & 13.25\% & 52  & 13.00\% \\
40 - 49 Years            & 337  & 18.72\% & 75  & 18.75\% & 68  & 17.00\% \\
50 - 59 Years            & 467  & 25.94\% & 89  & 22.25\% & 87  & 21.75\% \\
60 - 69 Years            & 341  & 18.94\% & 60  & 15.00\% & 77  & 19.25\% \\
70 - 75 Years            & 176  & 9.78\%  & 45  & 11.25\% & 39  & 9.75\%  \\ \hline
\multicolumn{7}{c}{Family status}                                         \\ \hline
Single                   & 185  & 10.28\% & 60  & 15.00\% & 49  & 12.25\% \\
Married                  & 1221 & 67.83\% & 259 & 64.75\% & 269 & 67.25\% \\
Other                    & 394  & 21.89\% & 81  & 20.25\% & 82  & 20.50\% \\ \hline
\multicolumn{7}{c}{Education level}                                       \\ \hline
No degree                & 3    & 0.17\%  & 3   & 0.75\%  & 2   & 0.50\%  \\
Secondary school diploma & 18   & 1.00\%  & 4   & 1.00\%  & 7   & 1.75\%  \\
Secondary school diploma & 202  & 11.22\% & 46  & 11.50\% & 46  & 11.50\% \\
Secondary school leav. certif. &
  539 &
  29.94\% &
  117 &
  29.25\% &
  115 &
  28.75\% \\
Technical college degree & 646  & 35.89\% & 143 & 35.75\% & 157 & 39.25\% \\
University degree        & 392  & 21.78\% & 87  & 21.75\% & 73  & 18.25\% \\ \hline
\multicolumn{7}{c}{Income level}                                          \\ \hline
less than 1,100 Euro     & 53   & 2.94\%  & 14  & 3.50\%  & 16  & 4.00\%  \\
1,101 - 1,300 Euro       & 33   & 1.83\%  & 6   & 1.50\%  & 6   & 1.50\%  \\
1,301 - 1,500 Euro       & 43   & 2.39\%  & 18  & 4.50\%  & 13  & 3.25\%  \\
1,501 - 1,700 Euro       & 53   & 2.94\%  & 19  & 4.75\%  & 4   & 1.00\%  \\
1,701 - 2,150 Euro       & 127  & 7.06\%  & 31  & 7.75\%  & 21  & 5.25\%  \\
2,151 - 2,600 Euro       & 177  & 9.83\%  & 41  & 10.25\% & 47  & 11.75\% \\
2,601 - 3,100 Euro       & 238  & 13.22\% & 47  & 11.75\% & 54  & 13.50\% \\
3,101 - 3,600 Euro       & 266  & 14.78\% & 43  & 10.75\% & 59  & 14.75\% \\
3,601 - 4,300 Euro       & 299  & 16.61\% & 58  & 14.50\% & 56  & 14.00\% \\
4,301 - 5,500 Euro       & 286  & 15.89\% & 60  & 15.00\% & 76  & 19.00\% \\
5,501 - 5,700 Euro       & 72   & 4.00\%  & 20  & 5.00\%  & 21  & 5.25\%  \\
5,701 - 6,400 Euro       & 56   & 3.11\%  & 11  & 2.75\%  & 9   & 2.25\%  \\
6,401 Euro or more       & 97   & 5.39\%  & 32  & 8.00\%  & 18  & 4.50\%  \\ \hline
\multicolumn{7}{c}{State affiliation}                                     \\\hline
Baden-Württemberg        & 219  & 12.17\% & 62  & 15.50\% & 49  & 12.25\% \\
Bavaria                  & 322  & 17.89\% & 72  & 18.00\% & 87  & 21.75\% \\
Berlin                   & 27   & 1.50\%  & 7   & 1.75\%  & 7   & 1.75\%  \\
Brandenburg              & 60   & 3.33\%  & 14  & 3.50\%  & 22  & 5.50\%  \\
Bremen                   & 12   & 0.67\%  & 1   & 0.25\%  & 5   & 1.25\%  \\
Hamburg                  & 25   & 1.39\%  & 7   & 1.75\%  & 7   & 1.75\%  \\
Hesse                    & 95   & 5.28\%  & 29  & 7.25\%  & 19  & 4.75\%  \\
Mecklenburg Western P    & 32   & 1.78\%  & 1   & 0.25\%  & 6   & 1.50\%  \\
Lower Saxony             & 230  & 12.78\% & 45  & 11.25\% & 37  & 9.25\%  \\
Northrhine-Westphalia    & 359  & 19.94\% & 74  & 18.50\% & 77  & 19.25\% \\
Rhineland Palatinate     & 118  & 6.56\%  & 28  & 7.00\%  & 20  & 5.00\%  \\
Saarland                 & 38   & 2.11\%  & 11  & 2.75\%  & 6   & 1.50\%  \\
Saxony                   & 74   & 4.11\%  & 16  & 4.00\%  & 13  & 3.25\%  \\
Saxony-Anhalt            & 45   & 2.50\%  & 10  & 2.50\%  & 18  & 4.50\%  \\
Schleswig Holstein       & 80   & 4.44\%  & 9   & 2.25\%  & 6   & 1.50\%  \\
Thuringia                & 64   & 3.56\%  & 14  & 3.50\%  & 21  & 5.25\% 
\end{longtable}
\normalsize

%Representativeness
\newpage
\section{Results}

\begin{table}[htp]
\scriptsize
\centering
\caption{Correlation of latent variables of the extended \acrshort{tpb} for \acrshort{pv}}
\label{tab:CorrPV}
\begin{tabular}{l|llllllll}
     & ENVC   & NOVS  & FINB  & ENVB  & ATT   & SOCN  & PBC   & INT   \\ \hline
ENVC & 1.000  &       &       &       &       &       &       &       \\
NOVS & -0.015 & 1.000 &       &       &       &       &       &       \\
FINB & 0.200  & 0.258 & 1.000 &       &       &       &       &       \\
ENVB & 0.377  & 0.217 & 0.545 & 1.000 &       &       &       &       \\
ATT  & 0.262  & 0.325 & 0.641 & 0.732 & 1.000 &       &       &       \\
SOCN & 0.187  & 0.413 & 0.548 & 0.571 & 0.703 & 1.000 &       &       \\
PBC  & 0.028  & 0.176 & 0.362 & 0.383 & 0.456 & 0.345 & 1.000 &       \\
INT  & 0.038  & 0.417 & 0.553 & 0.427 & 0.632 & 0.544 & 0.394 & 1.000
\end{tabular}
\end{table}

\begin{table}[htp]
\scriptsize
\centering
\caption{Correlation of latent variables of the extended \acrshort{tpb} for \acrshort{eea}}
\label{tab:CorrEEA}
\begin{tabular}{l|llllllll}
     & ENVC   & NOVS   & FINB   & ENVB   & ATT    & SOCN   & PBC    & INT   \\ \hline
ENVC & 1.000  &        &        &        &        &        &        &       \\
NOVS & -0.133 & 1.000  &        &        &        &        &        &       \\
FINB & 0.305  & 0.207  & 1.000  &        &        &        &        &       \\
ENVB & 0.456  & 0.236  & 0.603  & 1.000  &        &        &        &       \\
ATT  & 0.379  & 0.302  & 0.677  & 0.761  & 1.000  &        &        &       \\
SOCN & 0.195  & 0.374  & 0.541  & 0.608  & 0.676  & 1.000  &        &       \\
PBC  & 0.202  & -0.378 & -0.167 & -0.141 & -0.192 & -0.272 & 1.000  &       \\
INT  & 0.096  & 0.430  & 0.345  & 0.349  & 0.402  & 0.412  & -0.346 & 1.000
\end{tabular}
\end{table}

\begin{table}[]
\scriptsize
\centering
\caption{Correlation of latent variables of the extended \acrshort{tpb} for \acrshort{gt}}
\label{tab:CorrGT}
\begin{tabular}{l|llllllll}
     & ENVC   & NOVS  & FINB  & ENVB  & ATT   & SOCN  & PBC   & INT   \\ \hline
ENVC & 1.000  &       &       &       &       &       &       &       \\
NOVS & -0.078 & 1.000 &       &       &       &       &       &       \\
FINB & 0.264  & 0.110 & 1.000 &       &       &       &       &       \\
ENVB & 0.497  & 0.111 & 0.517 & 1.000 &       &       &       &       \\
ATT  & 0.443  & 0.188 & 0.555 & 0.803 & 1.000 &       &       &       \\
SOCN & 0.226  & 0.296 & 0.514 & 0.586 & 0.669 & 1.000 &       &       \\
PBC  & 0.246  & 0.134 & 0.440 & 0.499 & 0.473 & 0.410 & 1.000 &       \\
INT  & 0.238  & 0.348 & 0.522 & 0.549 & 0.662 & 0.563 & 0.400 & 1.000
\end{tabular}
\end{table}

\end{document}